\documentclass[11pt]{article}
\usepackage{amssymb,amsthm,amsmath,float,epsfig}
\usepackage[a4paper, includefoot, margin=0.5in]{geometry}
\usepackage{graphicx}
\usepackage{enumerate}
\usepackage{color}
\usepackage[affil-it]{authblk}
\usepackage{caption}
\usepackage{subcaption}
\usepackage[english]{babel}
\usepackage{titling}
\usepackage{multirow}
\usepackage{parskip}
\usepackage{hyperref}

\title{Characterizing interactions in online social networks \\during exceptional events}
\author{Elisa Omodei, Manlio De Domenico, and Alex Arenas}
\date{}
\affil{Department d'Enginyeria Inform\`{a}tica i Matem\`{a}tiques, Universitat Rovira i Virgili}

\begin{document}

\maketitle

\begin{abstract}
Nowadays, millions of people interact on a daily basis on online social media like Facebook and Twitter, where they share and discuss information about a wide variety of topics. In this paper, we focus on a specific online social network, Twitter, and we analyze multiple datasets each one consisting of individuals' online activity before, during and after an exceptional event in terms of volume of the communications registered. We consider important events that occurred in different arenas that range from policy to culture or science. For each dataset, the users' online activities are modeled by a multilayer network in which each layer conveys a different kind of interaction, specifically: retweeting, mentioning and replying. 
This representation allows us to unveil that these distinct types of interaction produce networks with different statistical properties, in particular concerning the degree distribution and the clustering structure. These results suggests that models of online activity cannot discard the information carried by this multilayer representation of the system, and should account for the different processes generated by the different kinds of interactions. Secondly, our analysis unveils the presence of statistical regularities among the different events, suggesting that the non-trivial topological patterns that we observe may represent universal features of the social dynamics on online social networks during exceptional events.\\
\\Keywords: multilayer, social networks, complex networks, exceptional events, big data 
\end{abstract}

\section{Introduction}

The advent of online social platforms and their usage in the last decade, with exponential increasing trend, made possible the analysis of human behavior with an unprecedented volume of data. To a certain extent, online interactions represent a good proxy for social interactions and, as a consequence, the possibility to track the activity of individuals in online social networks allows one to investigate human social dynamics~\cite{centola2010spread}.

More specifically, in the last years an increasing number of researchers focused on individual's activity in Twitter, a popular microblogging social platform with about 302 millions active users posting, daily, more than 500 millions messages (i.e., \emph{tweets}) in 33 languages\footnote{\url{https://about.twitter.com/company}}. In traditional social science research the size of the population under investigation is very small, with increasing costs in terms of human resources and funding. Conversely, monitoring Twitter activity, as well as other online social platforms as Facebook and Foursquare to cite just some of them, dramatically reduces such costs and allows to study a larger population sample, ranging from hundreds to millions of individuals~\cite{borge2013cascading}, within the emerging framework of computational social science~\cite{lazer2009life}.

The analysis of Twitter revealed that online social networks exhibit many features typical of social systems, with strongly clustered individuals within a scale-free topology~\cite{kwak2010twitter}. Twitter data~\cite{gonccalves2011modeling} has been used to validate Dunbar's theory about the theoretical cognitive limit on the number of stable social relationships~\cite{dunbar1992neocortex,dunbar20025}. It has been shown that individuals tend to share ties within the same metropolitan region and that non-local ties distance, borders and language differences affect their relationships~\cite{takhteyev2012geography}. Many studies were devoted to determine which and how information flows through the network~\cite{java2007we,yang2010predicting,wu2011says,myers2012information}, as well as to understand the mechanisms of information spreading -- e.g., as in the case of viral content -- to identify influential spreaders and comprehend their role~\cite{bakshy2011everyone,gonzalez2012assessing,borge2013emergence,gonzalez2013broadcasters,banos2013role}. Attention has also been given to investigate social dynamics during emergence of protests~\cite{borge2011structural}, with evidences of social influence and complex contagion providing an empirical test to the recruitment mechanisms theorized in formal models of collective action~\cite{gonzalez2011dynamics}.

Twitter allows users to communicate through small messages, using three different actions, namely mentioning, replying and retweeting. While some evidences have shown that users tend to exploit in different ways the actions made available by the Twitter platform~\cite{conover2011political}, such differences have not been quantified so far. In this work, we analyze the activities of users from a new perspective and focus our attention on how individuals interact during exceptional events. 

In our framework, an exceptional event is a circumstance not likely in everyday news, limited to a short amount of time -- typically ranging from hours to a few days -- that causes an exceptional volume of tweets, allowing to perform a significant statistical analysis of social dynamics. It is worth mentioning that fluctuations in the number of tweets, mentions, retweets and replies among users may vary from tens up to thousands in a few minutes, depending on the event. A typical example of exceptional event is provided by the discovery of the Higgs boson in July 2012~\cite{dedomenico2013anatomy}, one of the greatest events in modern physics.

We use empirical data collected during six exceptional events of different type, to shed light on individual dynamics in the online social network. We use social network analysis to quantify the differences between mentioning, replying and retweeting in Twitter and, intriguingly, our findings reveal universal features of such activities during exceptional events.

\section{Material \& Methods}

\subsection{Material}

It has been recently shown that the choice of how to gather Twitter data may significantly affect the results. In fact, data obtained from a simple backward search tend to over-represents more central users, not offering an accurate picture of peripheral activity, with more relevant bias for the network of mentions~\cite{gonzalez2012assessing}. Therefore, we used the streaming Application
Programming Interface (API) made available by Twitter, to collect all messages posted on the social network satisfying a set of temporal and semantic constraints. 

We consider different exceptional events because of their importance in different subjects, from politics to sport. More specifically, we focus on the Cannes Film Festival in 2013\footnote{\url{https://en.wikipedia.org/wiki/2013_Cannes_Film_Festival}} (Cannes2013), the discovery of the Higgs boson in 2012\footnote{\url{https://en.wikipedia.org/wiki/Higgs_boson\#Discovery_of_candidate_boson_at_CERN}}~\cite{dedomenico2013anatomy} (HiggsDiscovery2012), the 50th anniversary of Martin Luther King's famous public speech ``I have a dream'' in 2013\footnote{\url{https://en.wikipedia.org/wiki/I_Have_a_Dream}} (MLKing2013), the 14th IAAF World Championships in Athletics held in Moscow in 2013\footnote{\url{https://en.wikipedia.org/wiki/People's_Climate_March}} (MoscowAthletics2013), the ``People's Climate March'' -- a large-scale activist event to advocate global action against climate change -- held in New York in 2014\footnote{\url{https://en.wikipedia.org/wiki/People's_Climate_March}} (NYClimateMarch2014) and the official visit of US President Barack Obama in Israel in 2013\footnote{\url{https://en.wikipedia.org/wiki/List_of_presidential_trips_made_by_Barack_Obama\#2013}} (ObamaInIsrael2013).

For each event, we collected tweets sent between a starting time $t_{i}$ and a final time $t_{f}$ containing at least one keyword or hashtag, as specified in Table~\ref{tab:data}. It is worth remarking that in a few cases we complemented a dataset by including tweets obtained from the search API (at most 5\% of tweets with respect to the whole dataset).

\begin{table}[!b]
\centering
\begin{tabular}{| c || c | c | c |}
\hline
\textbf{Dataset} & \textbf{Starting date} & \textbf{Ending date}  & \textbf{Keywords}\\
\hline
\multirow{2}{*}{Cannes2013} & 06 May 2013 & 03 Jun 2013 & cannes film festival,cannes, canneslive \\ 
					    & 05:23:49 GMT & 03:48:26 GMT & \#cannes2013,\#festivalcannes, \#palmdor \\ 
\hline
\multirow{2}{*}{HiggsDiscovery2012} & 30 Jun 2012 & 10 Jul 2012 & lhc, cern, boson, higgs\\ 
							& 21:11:19 GMT & 20:59:56 GMT &  \\ 
\hline
\multirow{2}{*}{MLKing2013} & 25 Aug 2013 & 02 Sep 2013 & Martin Luther King \\ 
					    & 13:41:36 GMT & 08:16:21 GMT & \#ihaveadream \\ 
\hline
\multirow{2}{*}{MoscowAthletics2013} & 05 Aug 2013 & 19 Aug 2013 & mos2013com, moscow2013, mosca2013  \\ 
							  & 09:25:46 GMT & 12:35:21 GMT & moscu2013,\#athletics  \\ 
\hline
\multirow{2}{*}{NYClimateMarch2014} & 18 Sep 2014 & 22 Sep 2014 &peopleclimatemarch, peoplesclimate  \\ 
							  & 22:46:19 GMT & 04:56:25 GMT & marciaxilclima, climate2014  \\ 
\hline
\multirow{2}{*}{ObamaInIsrael2013} & 19 Mar 2013 & 03 Apr 2013 & obama, israel  \\ 
							& 15:56:29 GMT & 21:24:34 GMT & palestina, peace \\ 
\hline
\end{tabular}
\caption{Information about events used in this work. Note that starting and ending dates reported here consider only tweets where users perform a social action, i.e. tweets without mentions, replies or retweets are not considered.}
\label{tab:data}
\end{table}

\begin{figure}
\begin{subfigure}{.3\linewidth}
\centering
\includegraphics[scale=.25]{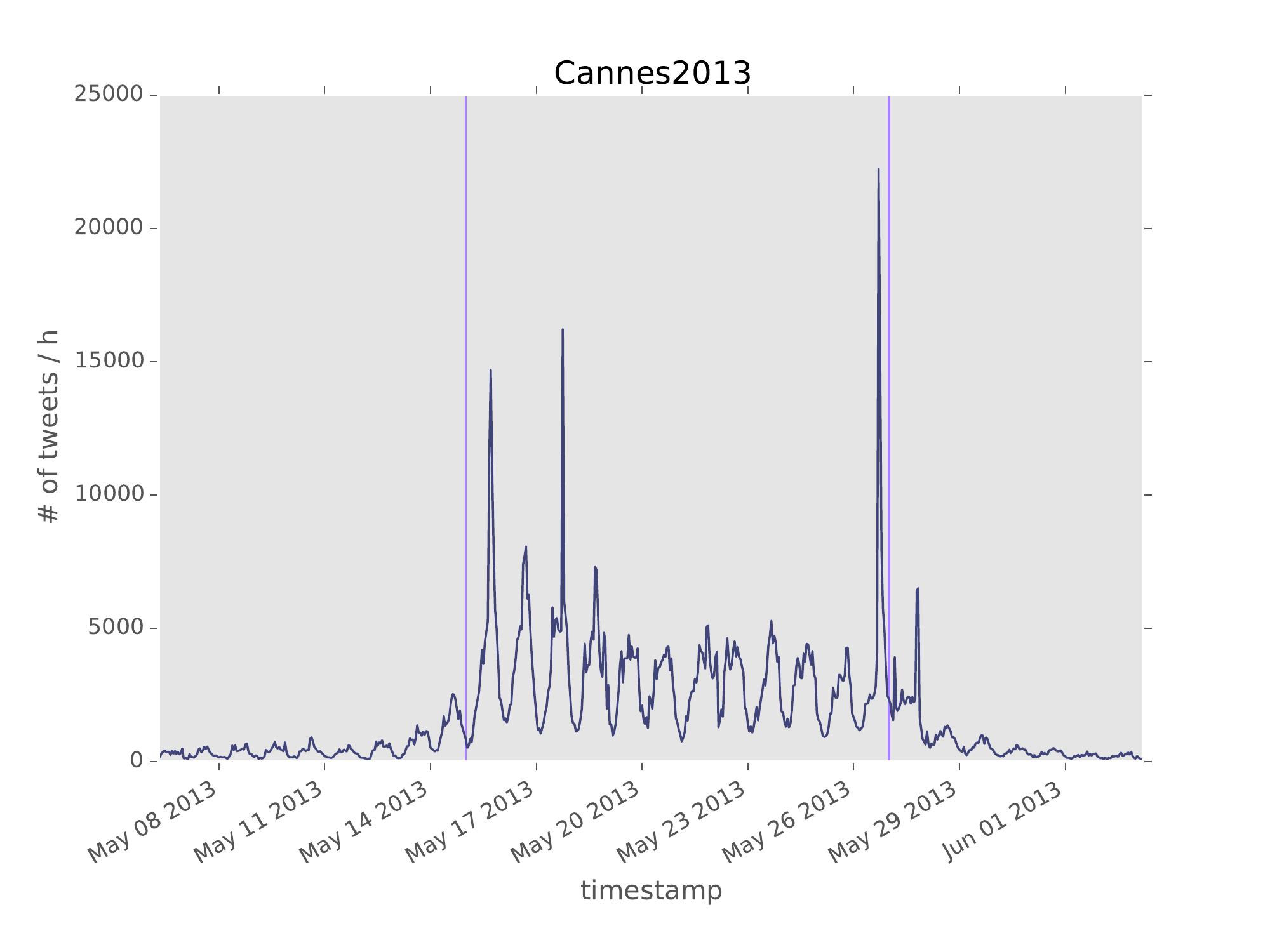}
\caption{}
\label{fig:sub1}
\end{subfigure}%
\begin{subfigure}{.3\linewidth}
\centering
\includegraphics[scale=.25]{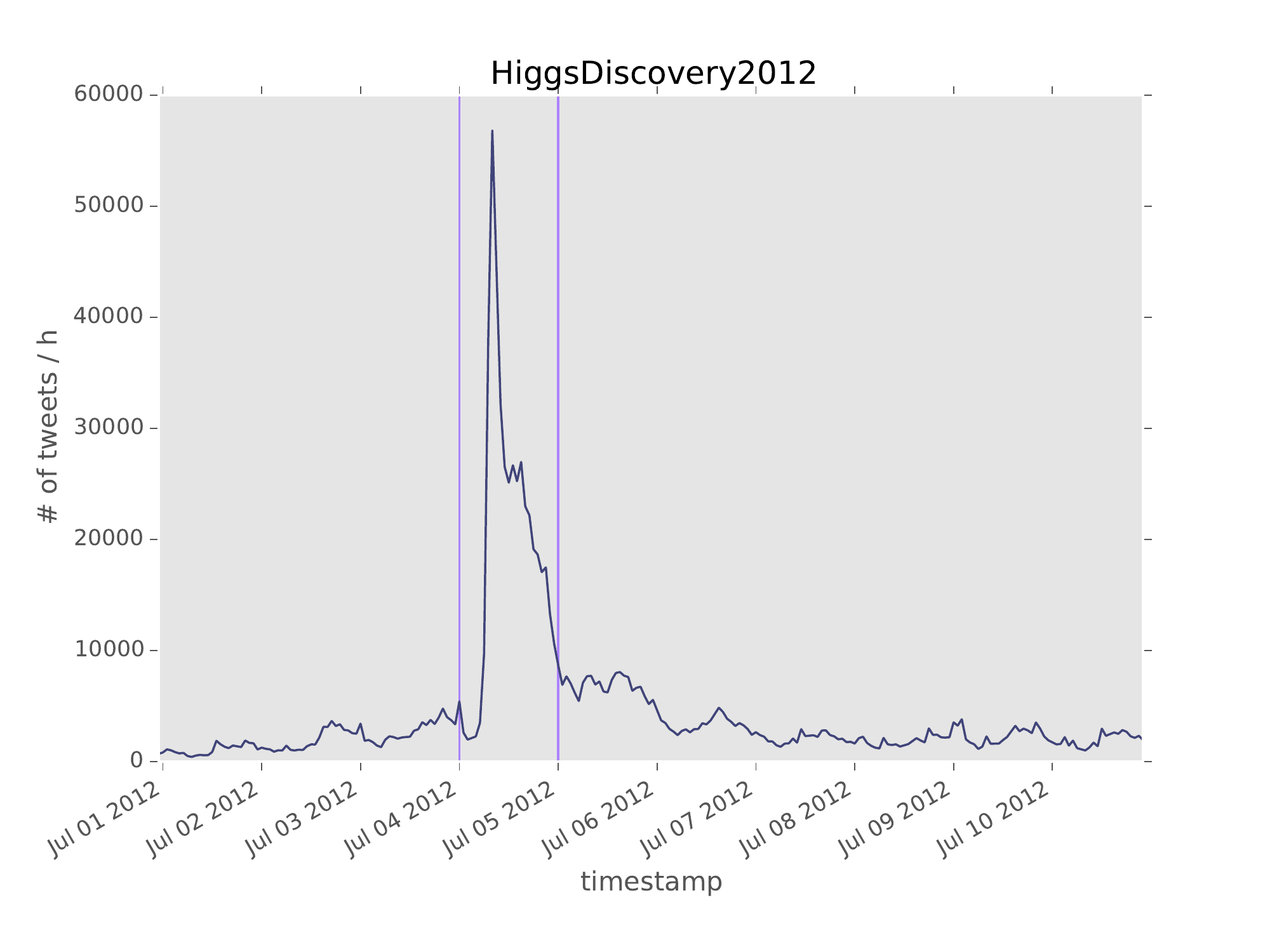}
\caption{}
\label{fig:sub2}
\end{subfigure}
\begin{subfigure}{.3\linewidth}
\centering
\includegraphics[scale=.25]{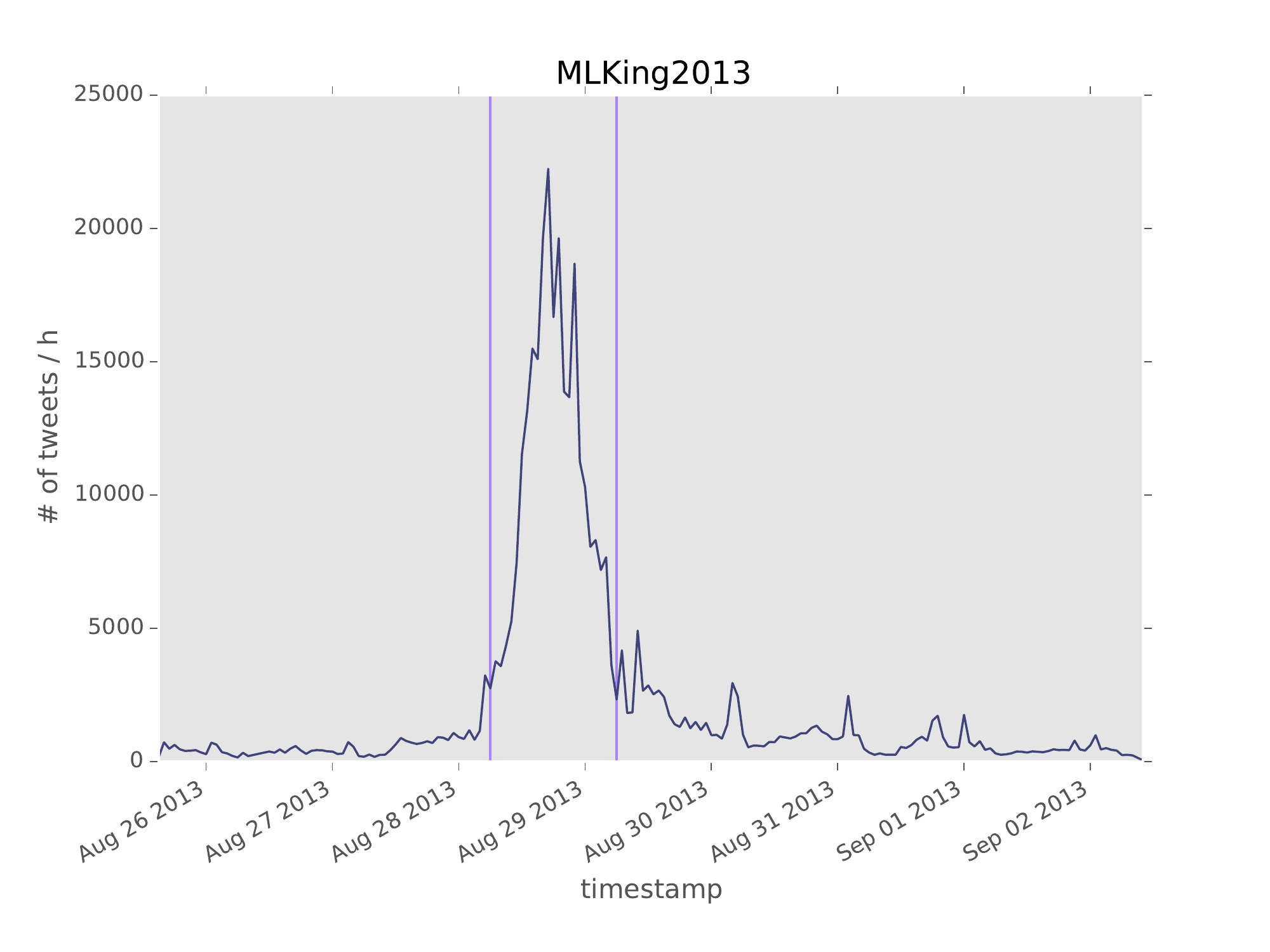}
\caption{}
\label{fig:sub3}
\end{subfigure}\\[1ex]
\begin{subfigure}{.3\linewidth}
\centering
\includegraphics[scale=.25]{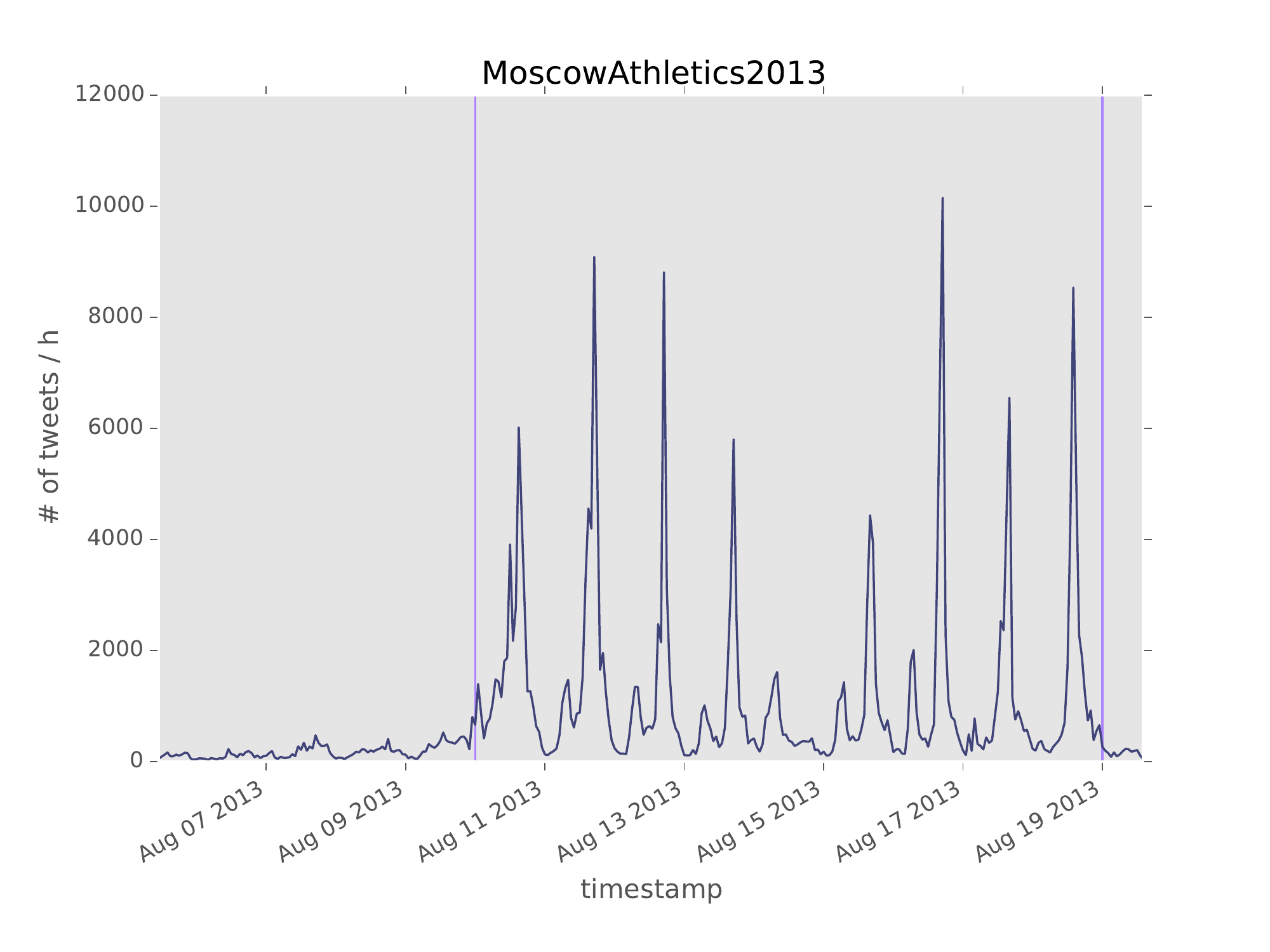}
\caption{}
\label{fig:sub1}
\end{subfigure}%
\begin{subfigure}{.3\linewidth}
\centering
\includegraphics[scale=.25]{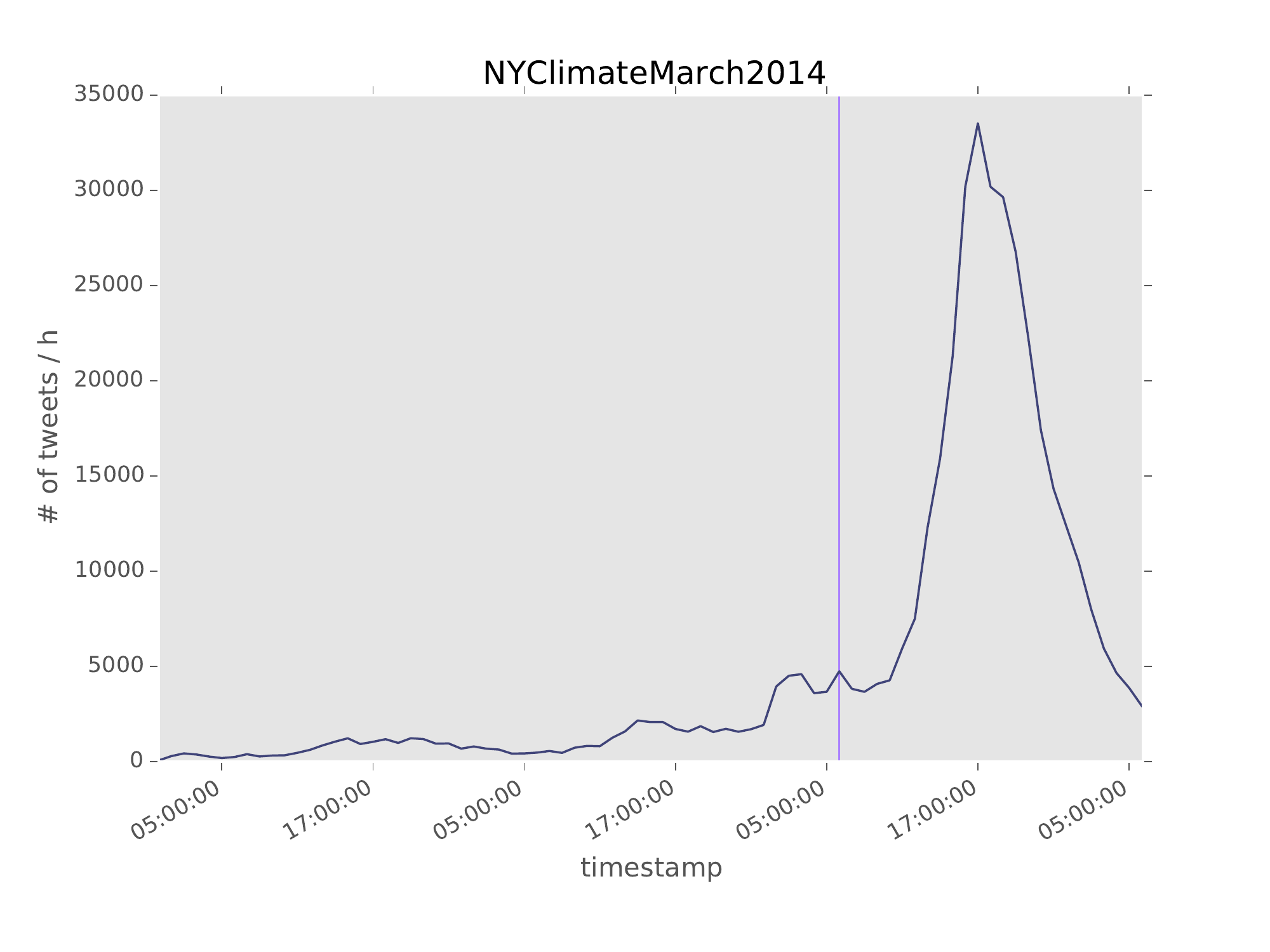}
\caption{}
\label{fig:sub2}
\end{subfigure}
\begin{subfigure}{.3\linewidth}
\centering
\includegraphics[scale=.25]{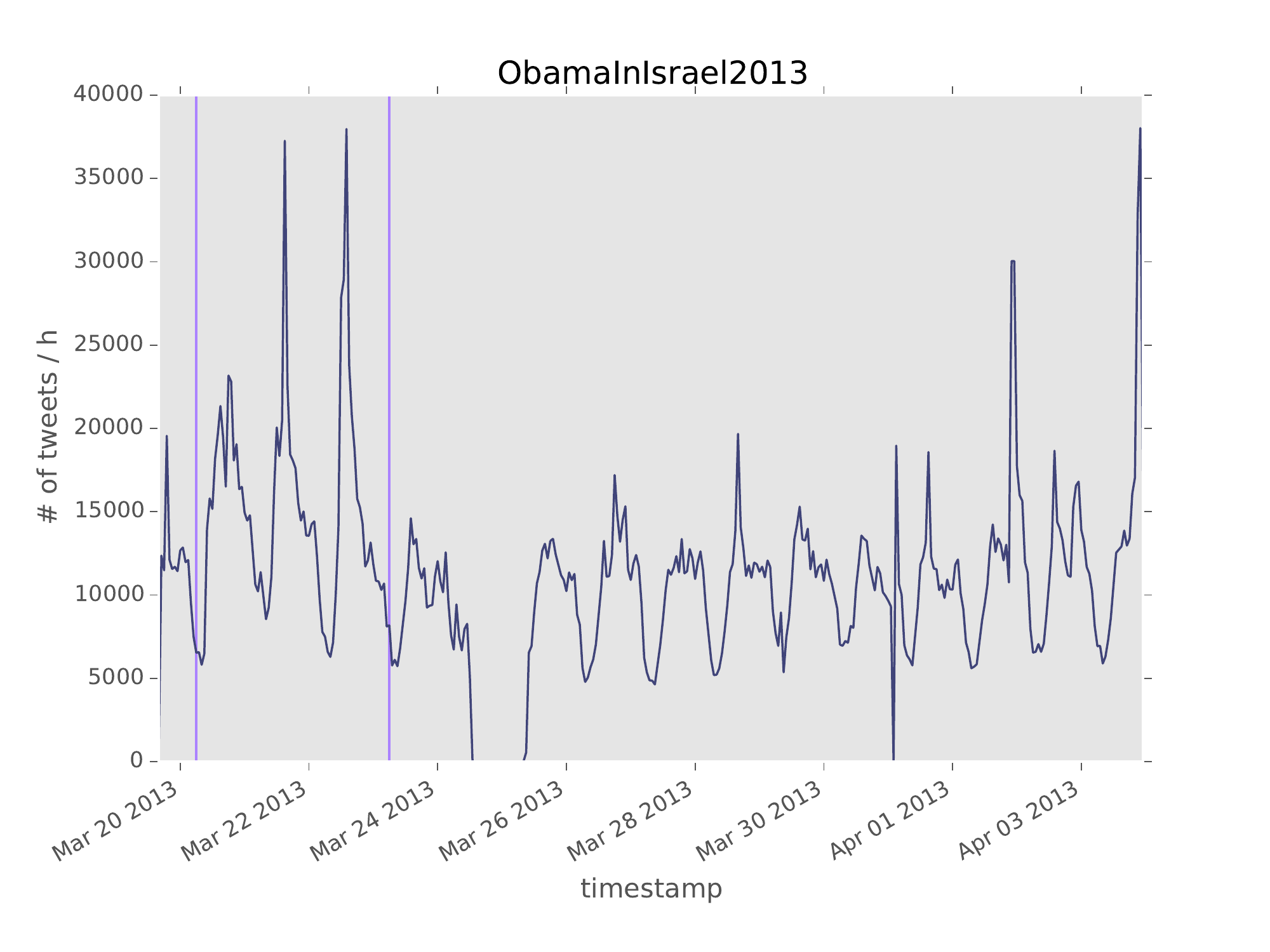}
\caption{}
\label{fig:sub3}
\end{subfigure}
\caption{Volume of tweets, in units of number of messages posted per hour, over time for the six exceptional events considered in our study.}
\label{tweets-volume}
\end{figure}

\subsection{Methods}

To understand the dynamics of Twitter user interactions during these exceptional events, we reconstruct, for each event, a network connecting users on the basis of the retweets, mentions and replies they have been the subject or object of. In the literature on Twitter data what is usually built is the network based on the follower-followee relationships between users \cite{kwak2010twitter,takhteyev2012geography,java2007we}. However, this kind of network only captures users' declared relations and it does not provide a good proxy for the actual interactions between them. Users, in fact, usually follow hundreds of accounts whose tweets appear in their news feed, even if there is no real interaction with the majority of those individuals. Therefore, to capture the social structure emerging from these interactions we build instead a network based on the exchanges between users, which can be deduced from the tweets that they produce. In particular, there are three kinds of interactions that can take place on Twitter and that we will focus on:
\begin{itemize}
\item A user can \textbf{retweet} (RT) another user's tweet. This means that the user is endorsing a piece of information shared by the other user, and is rebroadcasting it to her/his own followers.
\item A user can \textbf{reply} (RP) to another user's tweet. This represents an exchange from a user to another as a reaction of the information contained in a user's tweet.
\item A user can \textbf{mention} (MT) another user in a tweet. This represents an explicit share of a piece of information with the mentioned user.
\end{itemize}
A fourth kind of possible interaction is to favourite a user's tweet, which represents a simple endorsement of the information contained in the tweet, without rebroadcasting. However we do not have this kind of information for this dataset and therefore we do not consider this kind of interaction.

As just discussed, each kind of activity on Twitter (retweet, reply, and mention) represents a particular kind of interaction between two users. Therefore an appropriate framework to capture the overall structure of these interactions without loss of information about the different types is the framework of multilayer networks~\cite{cardillo2012emergence,dedomenico2013mathematical,nicosia2013growing,gallotti2013anatomy,dedomenico2015structural,dedomenico2015ranking}. More specifically, in the case under investigation the more appropriate model is given by edge-colored graphs, particular multilayer networks where a color is assigned to different relationships -- i.e., the edges -- among individuals defining as many layers as the number of colors. We refer to \cite{kivela2014multilayer} and \cite{boccaletti2014structure} for thorough reviews about multilayer networks.

\begin{table}[!b]
\centering
\begin{tabular}{| c || c | c | c | c |}
\hline
\textbf{Event} & \textbf{Aggregate} & \textbf{RT} & \textbf{RP} & \textbf{MT} \\
\hline
\multirow{2}{*}{Cannes2013} & N = 514,328 & 337,089 & 85,414 & 91,825 \\ 
& E = 700,492 & 490,268 & 82,952 & 127,272 \\ 
\hline
\multirow{2}{*}{HiggsDiscovery2012} & N = 747,659 & 434,687 & 167,385 & 145,587 \\ 
& E = 817,877 & 542,808 & 122,761 & 152,308 \\ 
\hline
\multirow{2}{*}{MLKing2013} & N = 346,069 & 286,227 & 24,664 & 35,178 \\ 
& E = 339,143 & 288,543 & 18,157 & 32,443 \\ 
\hline
\multirow{2}{*}{MoscowAthletics2013} & N = 103,319 & 73,377 & 11,983 & 17,959 \\ 
& E = 144,591 & 102,842 & 12,768 & 28,981 \\ 
\hline
\multirow{2}{*}{NYClimateMarch2014} & N = 115,284 & 94,300 & 7,900 & 13,084 \\ 
& E = 239,935 & 213,158 & 8,038 & 18,739 \\ 
\hline
\multirow{2}{*}{ObamaInIsrael2013} & N = 2,641,052 & 1,443,929 & 737,353 & 459,770 \\ 
& E = 2,926,777 & 1,807,160 & 586,074 & 533,543 \\ 
\hline
\end{tabular}
\caption{Number of nodes and edges of the network corresponding to each event considered in this study. The second column reports the total number of nodes and edges, corresponding to a network in which  information is aggregated. The last three columns report the number of active nodes and edges per layer. A node is considered active on a given layer if the corresponding user is the subject or the object of the corresponding kind of interaction.}
\label{network-stats}
\end{table}

\begin{figure}
\centering
\fbox{\includegraphics[width=\textwidth]{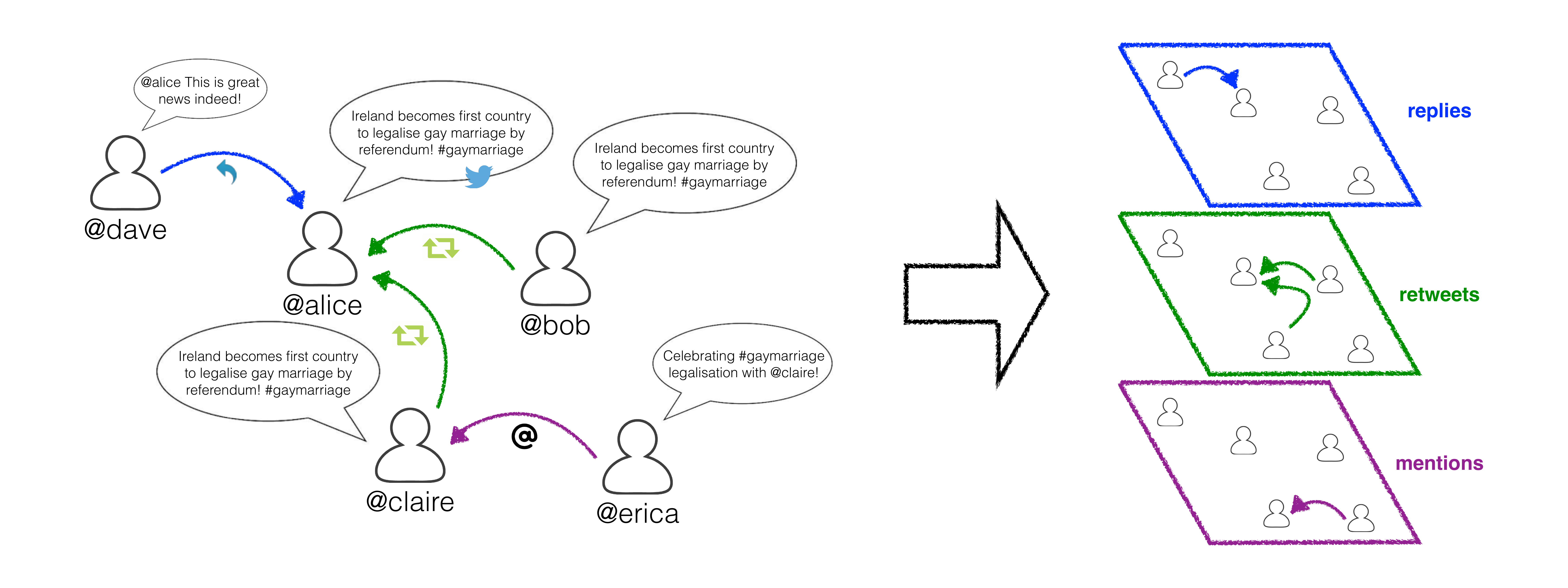}}
\caption{Illustrative example of a multilayer network representing the different interactions between Twitter users in the context of an exceptional event. Different colors are assigned to different actions.}
\label{twitter-activity-example}
\end{figure}

Here, for each event, we build a multilayer network composed by $L=3$ layers $\{$RT,RP,MT$\}$, corresponding to the three actions that users can perform in Twitter, and $N$ nodes, being $N$ the number of Twitter users interacting in the context of the given event. A directed edge between user $i$ and user $j$ on the RT layer is assigned if $i$ retweeted $j$. Similarly, an edge exists on RP layer if user $i$ replied to user $j$, and on MT layer if $i$ mentioned $j$. An illustrative example is shown in Figure~\ref{twitter-activity-example}. 

Details about the number of nodes and edges characterizing each event are reported in Table~\ref{network-stats}. We can observe that the number of nodes and edges can vary importantly across events and across layers, but for each event and each interaction type the size of the corresponding networks is sufficient to allow a statistically significant analysis of the data.

\section{Results}

In the following we present an analysis of the networks introduced in the previous section, which is oriented at exploring two different but complementary questions. 

Firstly we want to know if, within one same event, the three kinds of interactions produce different network topologies. To this aim, we consider basic multilayer and single-layer network descriptors relevant to characterize social relationships, and we study how they vary when considering different layers.

Secondly, we want to unveil if different exceptional events present any common pattern regarding users interactions. As shown in Figure~\ref{tweets-volume}, the temporal pattern of the different events considered in our study presents highly heterogeneous profiles. Some events are, in fact, limited to one day or only to a few hours, whereas others span over a week or more, and the profile of tweets volume varies accordingly. However, despite of these differences, do the user interactions that take place during these events present any common feature?

\subsection{Edge overlap across layers}

\begin{figure}[!t]
\centering
\includegraphics[width=0.75\textwidth]{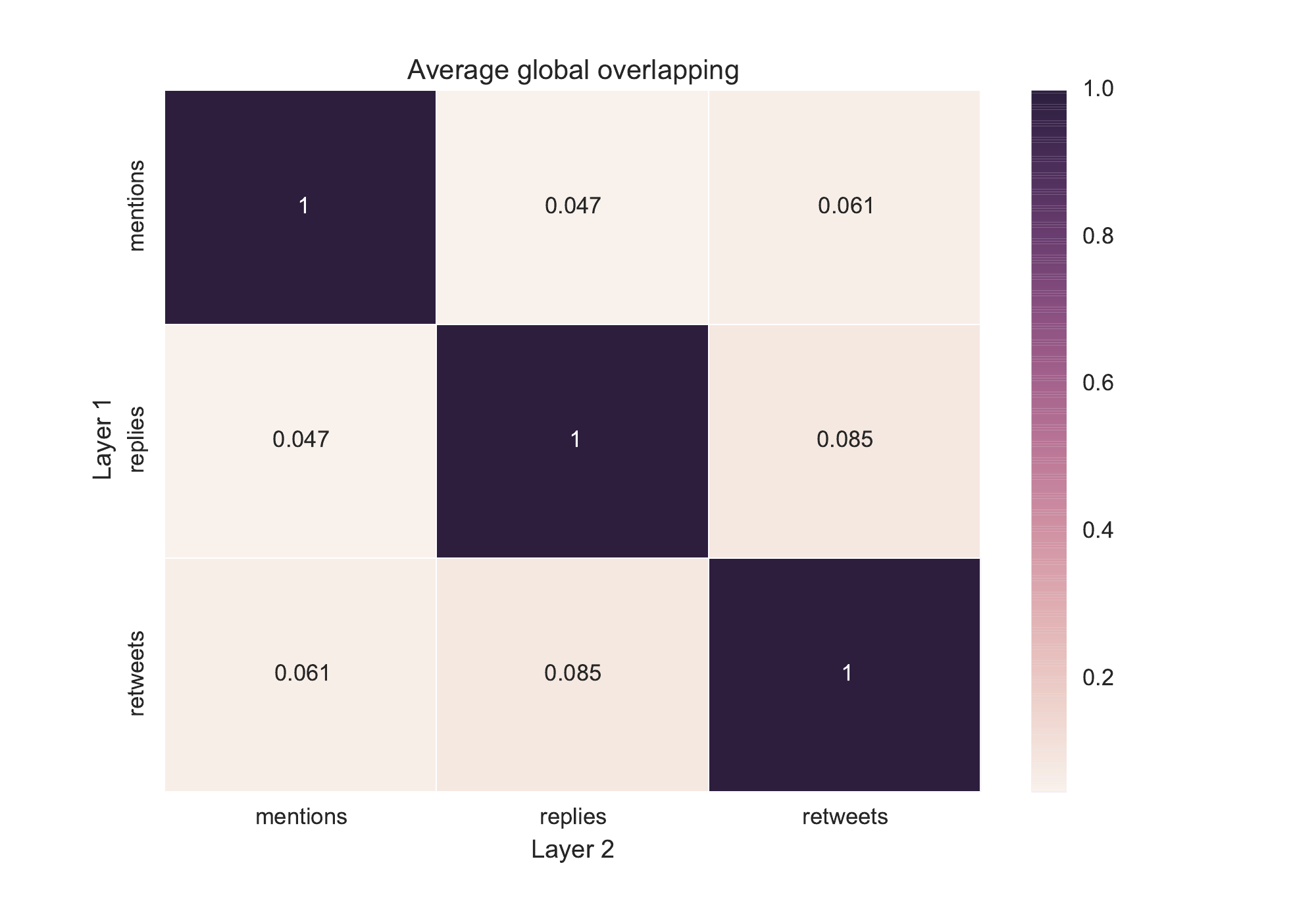}
\caption{Heatmap representing the edge overlap between pairs of layers, averaged over the different events.}
\label{edge-overlap}
\end{figure}

To understand if the kinds of interaction produce similar networks or not, we analyze if users interact similarly with each other regardless of the type of activity (retweet, reply or mention), or not. This information can be obtained by calculating the edge overlap \cite{PhysRevE.89.032804,dedomenico2015structural} between each pair of layers. However, when the number of edges is very heterogeneous across layers, a more suitable descriptor of edge overlap is given by
\begin{equation}
o_{\alpha\beta} = \frac{| E_\alpha \cap E_\beta |}{\mbox{min}(|E_\alpha|,|E_\beta |)},
\end{equation}
where $E_\alpha$ ($E_\beta$) is the set of edges belonging to layer $\alpha$ ($\beta$) and $|\cdot|$ indicates the cardinality of the set. This measure quantifies the proportion of pair-wise interactions -- represented by the edges -- that are common to two different layers. Because, as shown in Table~\ref{network-stats}, the number of edges can vary largely on the different layers, the normalization is given by the cardinality of the smallest set of edges, to avoid biases resulting from the size difference. 
The results are reported in Figure~\ref{edge-overlap}. Each value is obtained by averaging over the different events. The standard deviations are not shown in the figure for the sake of clarity, but are reported in Table~\ref{corr-table}. We see that, for every couple of layers, $(\alpha,\beta)$, $o_{\alpha\beta} \ll 1$. This result indicates that different layers contain different pairwise interactions, \textit{i.e.} the users that we retweet are not necessarily the same that we mention or we reply to, for example. This result suggests that considering the different activities separately might be very relevant in order to understand human interaction dynamics on Twitter.

\begin{table}[!b]
\centering
\begin{tabular}{| c | c | c |}
\hline
Layer pair & Edge overlap & Degree-degree correlation \\
\hline
MT-RP & $0.05 \pm 0.04$ & $0.50 \pm 0.12$ \\
MT-RT & $0.06 \pm 0.03$ & $0.33 \pm 0.08$ \\
RP-RT & $0.08 \pm 0.04$ & $0.35 \pm 0.10$ \\
\hline
\end{tabular}
\caption{Average and standard deviation across the different events of the edge overlap and of the degree-degree correlation, for each layer pair.}
\label{corr-table}
\end{table}

\subsection{Degree-degree correlations across layers}

\begin{figure}
\centering
\includegraphics[width=0.75\textwidth]{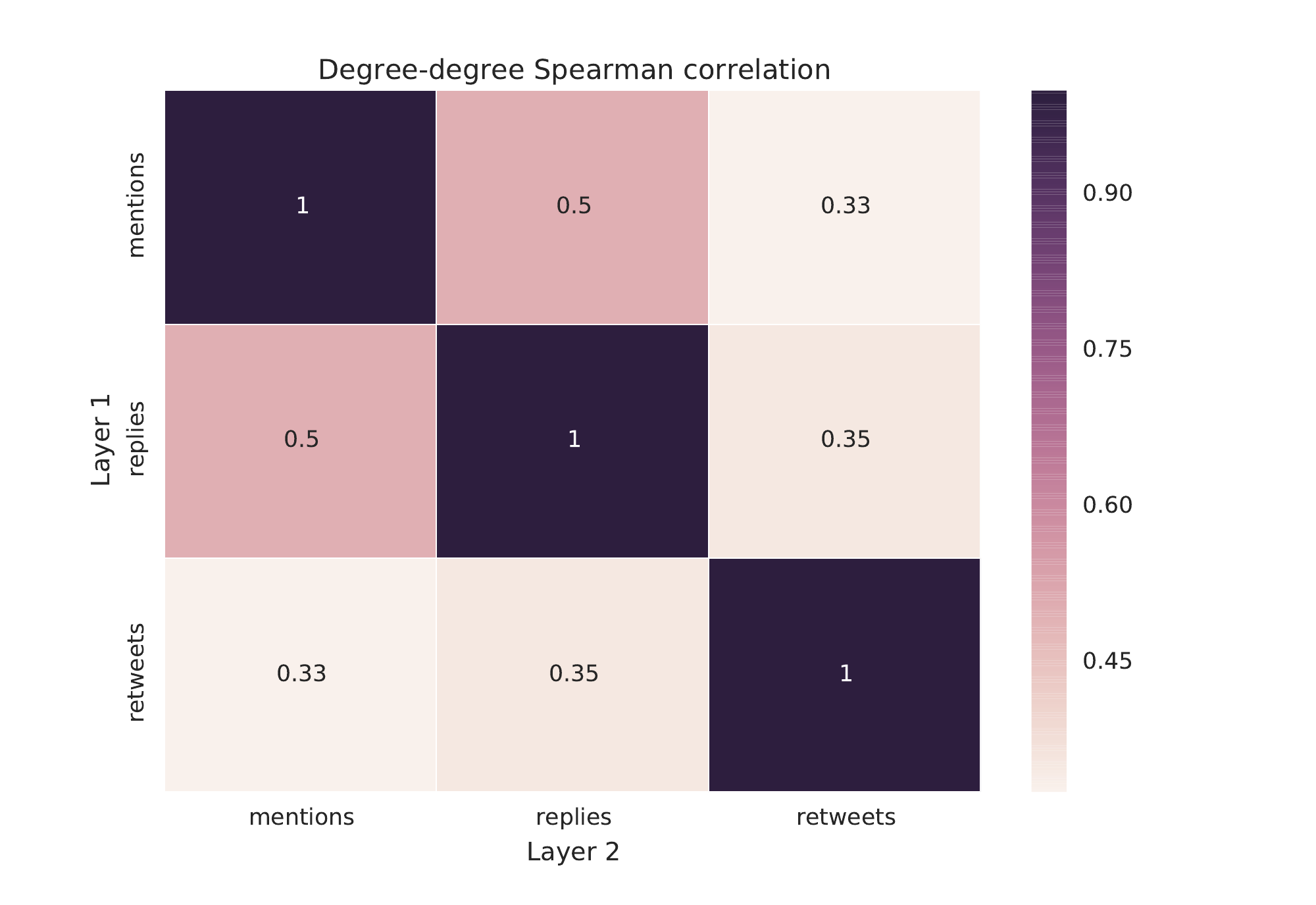}
\caption{Heatmap representing the average degree-degree correlation between layer pairs.}
\label{degree-degree-correlation}
\end{figure}

In this section, we study the degree connectivity of users, the most widely studied descriptor of the structure of a network. We focus in particular on the in-degree $k_{i,\alpha}$, which quantifies the number of users who interacted with user $i$ on layer $\alpha$ ($\alpha=$~RT, RP and MT). This is the simplest measure of the importance of the user in the network.

First, we explore if users have the same connectivity on the different layers, or not, \textit{i.e.} if the users consistently have the same degree of importance on all the layers, or not. To this aim, we compute the Spearman's rank correlation coefficient \cite{spearman1904proof} between the in-degree of users on one layer and their in-degree on a different layer, for each pair of layers. The results, averaged across the different events, are reported in Figure~\ref{degree-degree-correlation}, with statistical details reported in Table~\ref{corr-table}. The value of two degree-degree correlations out of three is about 0.35, and the third -- and highest -- correlation is 0.5. This means that users tend to have different in-degree values on the different layers, \textit{i.e.} a highly retweeted user is most likely not to be mentioned or replied to by as many users. This result represents a second important indicator that the different types of interaction produce different networks and should be considered separately in realistic modeling of individual dynamics.




\subsection{Degree distribution per layer}

\begin{figure}[!t]
\centering
\includegraphics[width=0.7\textwidth,angle=270]{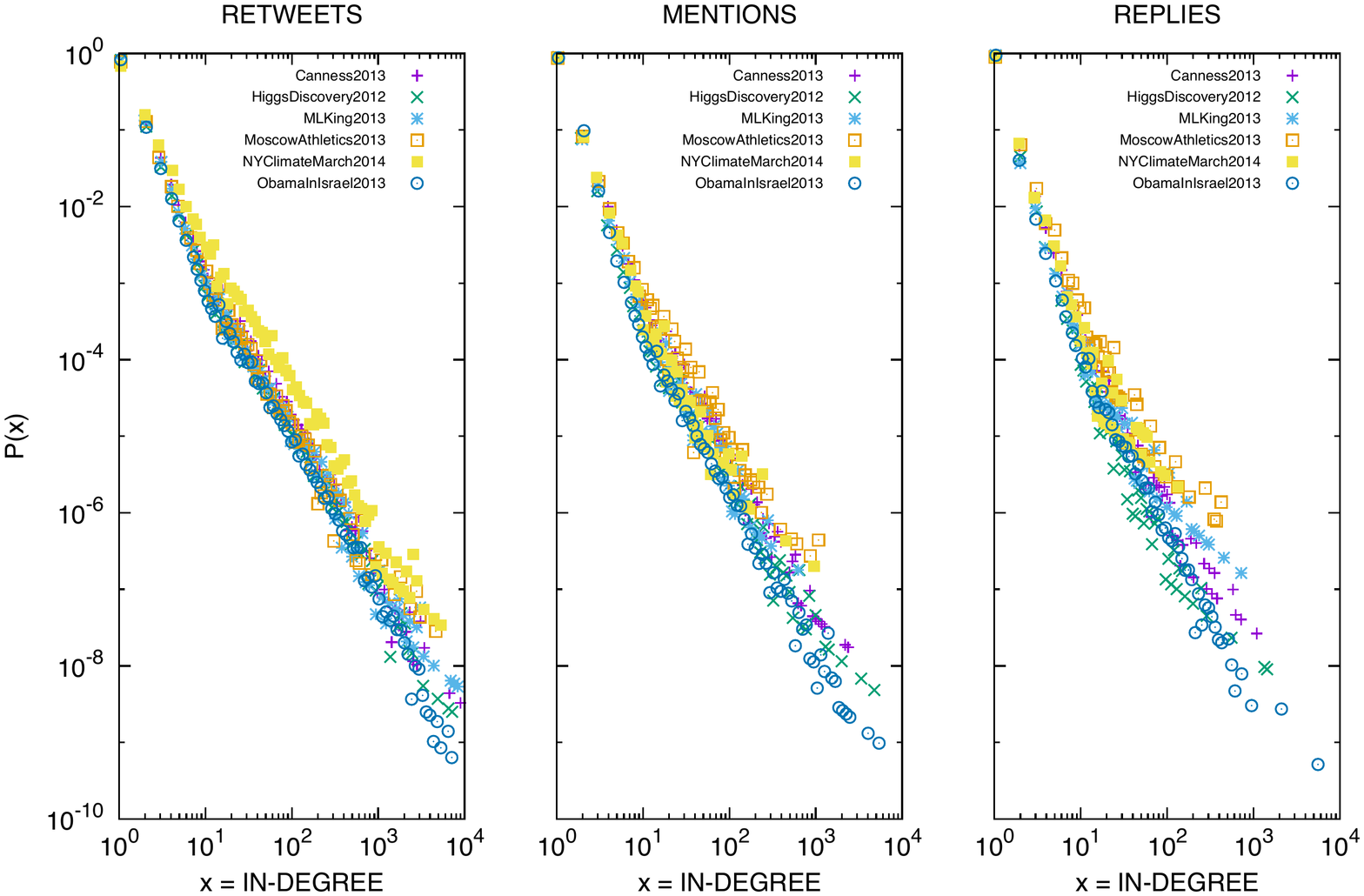}
\caption{Distribution of the in-degree for each event considered in this study (encoded by points with different shape and color) and each layer: retweets (left), mentions (center), and replies (right).}
\label{degree-distributions}
\end{figure}

Building on the result discussed in the previous section, we also explore, for each event, the distribution of the in-degree on the different layers, separately. Intriguingly, for each layer, we find that the empirical distributions corresponding to the all exceptional events present very similar shape, as shown in Figure~\ref{degree-distributions}. This result suggests that individuals' communications on Twitter present some universal characteristics across very different types of events .

The in-degree, shown in Figure~\ref{degree-distributions}, exhibits a power-law distribution for about three order of magnitudes. To validate our observation, we fit a power law to each distribution following a methodology similar to the one introduced in \cite{clauset2009power}. By noticing that the in-degree is a discrete variable, we estimate the scaling exponent of a discrete power law for each empirical distribution. The goodness of fit is estimated by using the Chi Square test \cite{snedecor1967statistical}. We find that the null hypothesis that the data is described by a discrete power law is accepted for all empirical distributions with a confidence level of $99\%$. We have tested other hypotheses, by considering other distributions with fat tails such as lognormal, exponential, Gumbel's extreme values, and Poisson. In the cases where the null hypothesis is accepted with the same confidence level, we used the Akaike information criterion (AIC) \cite{akaike1973information,akaike1974new} to select the best model. It is worth remarking that, in all cases, we find that the power law provide the best description of the data.

\begin{figure}
\centering
\includegraphics[width=0.75\textwidth]{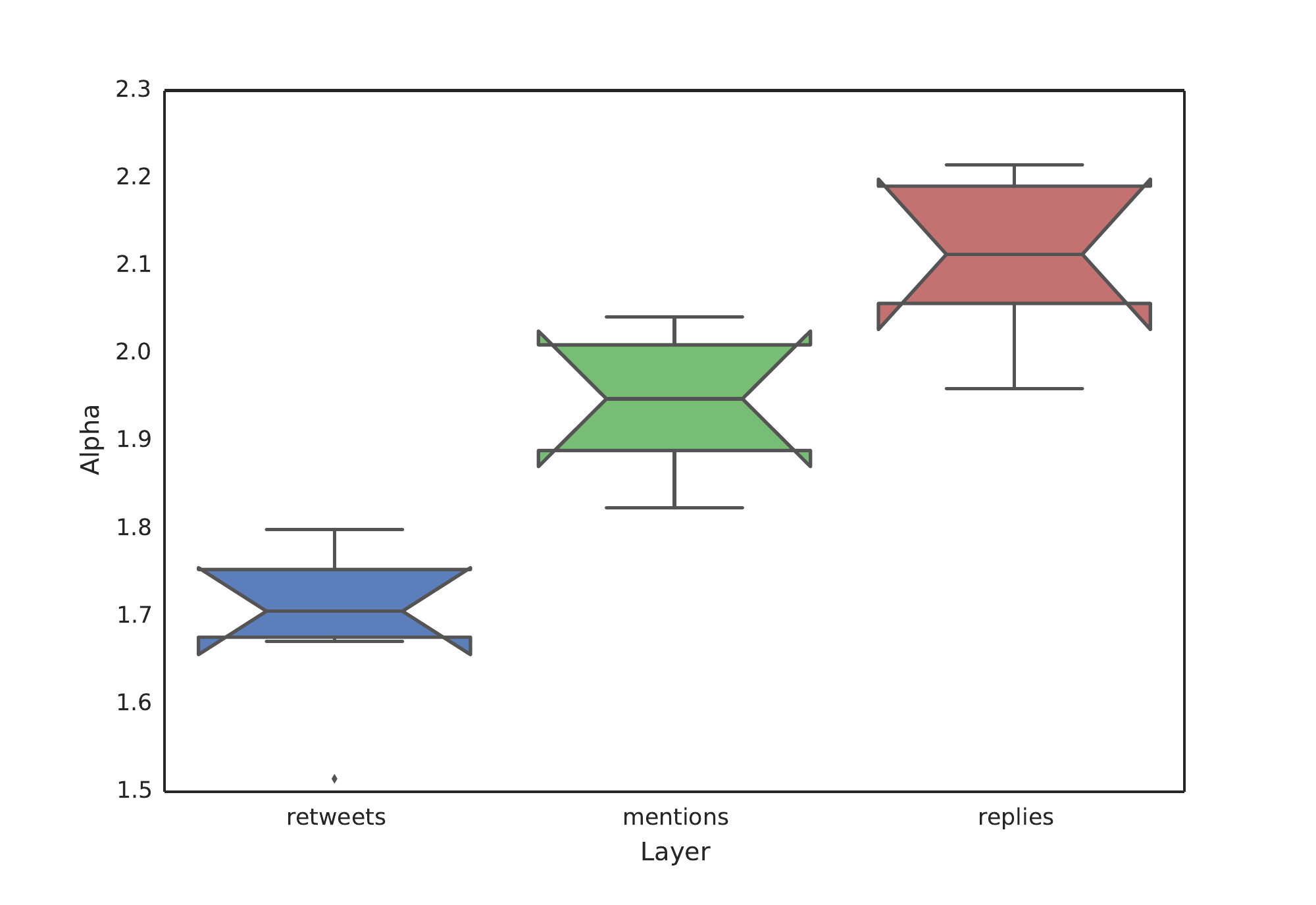}
\caption{Notched box plots showing the value of the scaling exponent of the in-degree distribution for each layer. Each box aggregates the values corresponding to the different events considered. Notched box plots present a contraction around the median, whose height is statistically important: if the notches of two boxes do not overlap, this offers evidence of a statistically significant difference between the two medians. This is the case here, meaning that the median scaling exponent of the in-degree distribution of each of the three layer is different from the exponent characterizing the in-degree distribution of the other layers.}
\label{alpha-notched-plots}
\end{figure}

Power-law distributions of the degree have been found in a large variety of empirical social networks \cite{caldarelli2007scale}. Here, the main finding of our results is that each kind of interaction presents a different scaling exponent. To show this, in Figure~\ref{alpha-notched-plots} we report three notched box plots, each corresponding to a different layer and including the information about the different events. Notched box plots present a contraction around the median, whose height is statistically important: if the notches of two boxes do not overlap, this offers evidence of a statistically significant difference between the two medians. This is indeed the case in Figure~\ref{alpha-notched-plots}, meaning that the median scaling exponent of the in-degree distribution of each of the three layer is different from the exponent characterizing the in-degree distribution of the other layers. The fact that the in-degree distributions corresponding to the different types of interaction are characterized by different scaling exponents indicates that the dynamics of each type of interaction in Twitter should be modeled as a distinct process, and that existing models of Twitter activity that do not take into account this fact should be carefully rethought.

\subsection{Average clustering per layer}

\begin{figure}
\centering
\includegraphics[width=0.75\textwidth]{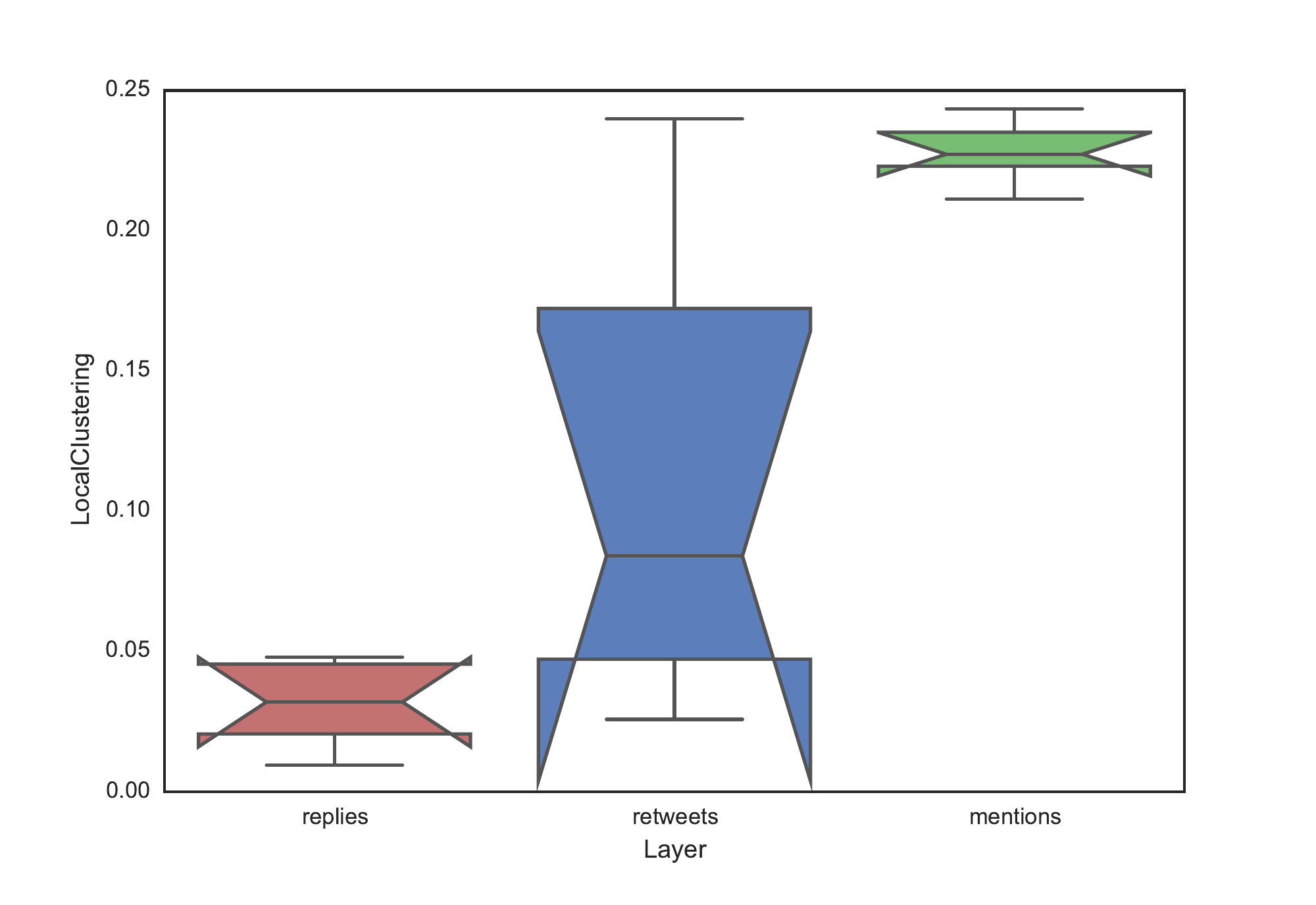}
\caption{Notched box plots showing the value of the average clustering coefficient for each layer. Each box aggregates the values corresponding to the different events considered.} 
\label{clustering-notched-plots}
\end{figure}

Lastly, for each layer separately, we calculate the average clustering coefficient of the corresponding network. This is a measure of the transitivity of the observed interactions, and constitutes an important metric to characterize social networks \cite{watts1998collective}. In particular, for each event and each layer, we compute the average local clustering coefficient defined by
\begin{equation}
\bar{C} = \frac{1}{N}\sum_{i=1}^{N} C_i,
\end{equation}
where
\begin{equation}
C_i = \frac{2|\{e_{jk}: v_j,v_k \in N_i, e_{jk} \in E\}|}{k_i(k_i-1)} \mbox{,}
\end{equation}
where $e_{jk}$ indicates the edge between users $j$ and $k$. We show in Figure~\ref{clustering-notched-plots} the values of the clustering coefficient using three notched box plots, each corresponding to a different layer and including the information about the different events. The mention network has the highest clustering level, whereas the reply network has the lowest one. The clustering level of the retweet network is the most variable across events, however the three medians are again different because the notches do not overlap. This result is a further confirmation that the three layers, and therefore the three types of interaction that they represent, form different network topologies and that the dynamical processes producing them are thus distinct.

\section{Discussion}

In this paper we analyze six datasets consisting of Twitter conversations surrounding distinct exceptional events. The considered events span over very different topics: entertainment, science, commemorations, sports, activism, and politics. Our results show that, despite the different fluctuations in time and in volume, there are some statistical regularities across the different events. In particular, we find that the in-degree distribution of users and the clustering coefficient in each of the three layers (representing interactions based on retweet, replies, and mentions, respectively) are the same across the six different events. Our first conclusion is therefore that users behavior on Twitter -- during exceptional events -- presents some universal patterns.

Secondly, we show that different types of interactions between users on Twitter (retweeting, replying and mentioning) generate networks presenting different topological characteristics. These differences were captured making use of the multilayer network framework: instead of discarding the information contained in the tweets regarding how users interact, we use this information to build a more complete representation of the system by means of three layers, each representing a different type of interaction. The fact that networks corresponding to different layer present different statistical properties is an important hint for models aiming at reproducing human behavior in online social networks. Our results indicate that, to faithfully represent how users interact, these models cannot be based on an aggregated view of the network and should account for all the different processes taking place in the system, separately.

\newpage
\bibliographystyle{unsrt}
\bibliography{paper_merge_arxiv}

\end{document}